\documentclass[a4paper,english,aps,prb,twocolumn,showpacs]{revtex4}
\usepackage[T1]{fontenc}
\usepackage[latin1]{inputenc}
\usepackage{amsmath}
\usepackage[dvips]{graphicx}
\usepackage{amssymb}

\makeatletter
\usepackage{color}
\usepackage{amsfonts}
\usepackage{amsmath}
\usepackage{amssymb}
\usepackage{graphicx}%
\usepackage{babel}
\makeatother
\begin{document}

\title{Possible origin of 60-K plateau in the YBa$_{2}$Cu$_{3}$O$_{6+y}$
phase diagram}

\author{T. A. Zaleski}

\affiliation{Department of Physics, University of Illinois at Urbana-Champaign,
Urbana, Illinois 61801, USA}

\affiliation{Institute of Low Temperature and Structure Research, Polish Academy
of Sciences P.O. Box 1410, 50-950 Wroc\l aw, Poland}

\author{T. K. Kope\'{c}}

\affiliation{Institute of Low Temperature and Structure Research, Polish Academy
of Sciences P.O. Box 1410, 50-950 Wroc\l aw, Poland}

\begin{abstract}
We study a model of YBa$_{2}$Cu$_{3}$O$_{6+y}$ to investigate the
influence of oxygen ordering and doping imbalance on the critical
temperature $T_{c}\left(y\right)$ and to elucidate a possible origin
of well-known feature of YBCO phase diagram: the 60-K plateau. Focusing
on ``phase only'' description of the high-temperature superconducting
system in terms of collective variables we utilize a three-dimensional
semi microscopic XY model with two-component vectors that involve
phase variables and adjustable parameters representing microscopic
phase stiffnesses. The model captures characteristic energy scales
present in YBCO and allows for strong anisotropy within basal planes
to simulate oxygen ordering. Applying spherical closure relation we
have solved the phase XY model with the help of transfer matrix method
and calculated $T_{c}$ for chosen system parameters. Furthermore,
we investigate the influence of oxygen ordering and doping imbalance
on the shape of YBCO phase diagram. We find it unlikely that oxygen
ordering alone can be responsible for the existence of 60-K plateau.
Relying on experimental data unveiling that oxygen doping of YBCO
may introduce significant charge imbalance between CuO$_{2}$ planes
and other sites, we show that \emph{simultaneously} the former are
underdoped, while the latter -- strongly overdoped almost in the whole
region of oxygen doping in which YBCO is superconducting. As a result,
while oxygen content is increased, this provides two counter acting
factors, which possibly lead to rise of 60K plateau. Additionally,
our result can provide an important contribution to understanding
of experimental data supporting existence of multicomponent superconductivity
in YBCO.
\end{abstract}


\maketitle

\section{Introduction}

\begin{figure}
\includegraphics[%
  scale=0.4]{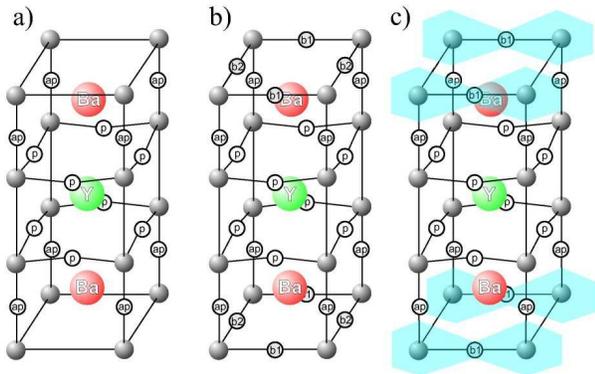}

\caption{(Color online) Crystal structure of YBa$_{2}$Cu$_{3}$O$_{6+y}$.
Grey and open circles are copper and oxide atoms, respectively. Additionally,
oxygen atoms are divided into three groups: from fully oxidized copper-oxide
layers (O$_{p}$), apical (O$_{ap}$), from oxygen deficient layers
(O$_{b1}$, O$_{b2}$). Effect on oxygen doping is also shown: a)
$y=0$ -- O$_{b1}$ and O$_{b2}$ sites are empty; b) $y\in\left(0,1\right)$
-- O$_{b1}$ and O$_{b2}$ sites occupied in random or ordered manner,
depending on actual value of $y$; c) $y=1$ -- oxygen is ordered
forming chains: all O$_{b1}$ sites are occupied and O$_{b2}$ sites
are empty.\label{cap:YBCO_structure}}
\end{figure}

The YB$_{2}$Cu$_{3}$O$_{6+y}$ compound (YBCO), as discovered in
1987 by Wu and co-workers, is the first material that became superconducting
in boiling nitrogen temperature.\cite{YBCO_discovery} The material
contains three copper-oxide layers in a unit cell: two of them are
separated by Yttrium atom, the third -- basal plane, is surrounded
by two Barium atoms (see, Fig. \ref{cap:YBCO_structure}). The oxygen
can be introduced into the basal plane and its content can be varied
from 6 to 7 per formula ($0\le y\le1$).\cite{YBCO_review} For $y=0$,
all O$_{b1}$ and O$_{b2}$ sites are empty and the system is tetragonal.
As the oxygen content increases, additional atoms occupy O$_{b1}$
and O$_{b2}$ sites randomly up to a critical doping, for which tetragonal-orthorhombic
(T-O) phase transition occurs. For higher dopings oxygen in O$_{b1}$
and O$_{b2}$ sites becomes partially ordered forming (depending on
oxygen content) one of three different orthorhombic phases: ortho-I
(with fragments of copper-oxide chains in the basal plane); ortho-II
(with chains in every second O$_{b1}$ site); and ortho-III (with
alternating one empty and two filled chains). Finally for $y=1$ (maximum
oxygen content), all O$_{b1}$ sites become occupied and all O$_{b2}$
-- empty. The oxygen in the basal plane acts as a charge reservoir
introducing holes into CuO$_{2}$ plane copper atoms. Charge concentration
can be also changed by substituting Yttrium atoms with Calcium, however
it seams that holes introduced in such way tend to remain in CuO$_{2}$
layers.\cite{Apical_oxygen} Amount of oxygen also determines electronic
state of the system. For $y<0.4$ the material is insulating, while
for $y>0.4$ becomes superconducting. The critical temperature is
very sensitive to oxygen doping and the temperature-oxygen amount
phase diagram contains two characteristic plateaus at 60K and 90K.
While the latter is now interpreted as an optimum doping with small
overdoping region, the origin of the 60K plateau is still not fully
clear. It was argued that it might be explained by ordering of the
oxygen atoms within the basal plane.\cite{O_ordering1,O_ordering2,O_ordering3,O_ordering4}
On the other hand, it was also suggested that the reason might be
purely electronic: superconductivity is weakened at the carrier concentration
of 1/8 leading to the plateau.\cite{YBCO_transport} It is a goal
of the present paper to investigate the influence of oxygen ordering
and doping imbalance on the critical temperature of YBCO using a model
that can accommodate strong anisotropy and characteristic energy scales
in order to determine a possible origin of 60-K plateau on the phase
diagram. 

Binding of electrons into pairs is essential in forming the superconducting
state, however its remarkable properties--zero resistance and Meissner
effect--require \emph{phase coherence} among the pairs as well. While
the phase order is unimportant for determining the value of the transition
temperature $T_{c}$ in conventional BCS superconductors, in materials
with low carrier density such as high-T$_{c}$ oxide superconductors,
phase fluctuations may have a profound influence on low temperature
properties.\cite{Phase_fluctuations} Measurements of the frequency-dependent
conductivity, in the frequency range 100-600 GHz, show that phase
correlations indeed persist above $T_{c}$, where the phase dynamics
is governed by the bare microscopic phase stiffnesses.\cite{Phase_stiffnesses}
As a result, for underdoped cuprate superconductors, the conventional
ordering of binding and phase stiffness energies appears to be reversed.
Thus, the issue how phase correlations develop is a central problem
of high-$T_{c}$ superconductivity. Furthermore, it was directly shown
that integrating the electronic degrees of freedom out in $t$-$t^{'}$-$U$-$J$
Hubbard model, which is believed to correctly describe strongly interacting
systems, leads to phase-only description of superconductivity.\cite{t_t_U_J}

In the present paper, we propose a semi-microscopic model of YBCO,
which is founded on microscopic phase stiffnesses that set the characteristic
energy scales: in-plane $J_{\|}$ and inter-plane $J_{\bot}$ couplings
of CuO$_{2}$ layers, in-plane $J_{\|}^{'}$ coupling of basal planes
and inter-plane $J_{\bot}^{'}$ coupling between neighboring basal
and CuO$_{2}$ planes. Additionally, the model contains a parameter
$\eta$, which controls anisotropy of $J_{\|}^{'}$ gradually turning
off basal in-plane coupling along $b$ direction while $\eta$ changes
from 1 to 0. Using our previous results,\cite{OurPrevious} we model
values of in-plane phase stiffnesses as a function of oxygen amount
to reproduce YBCO phase diagram and to elucidate the origin of the
60-K plateau. Our approach goes beyond the mean field level and is
able to capture both the effects of phase fluctuations and huge anisotropy
on the superconducting phase transition. 

The outline of the reminder of the paper is as follows. In Section
II we construct an anisotropic three-dimensional XY model. Furthermore,
we solve it in the spherical approximation with the help of the transfer
matrix method and obtain a dependence of the critical temperature
on model parameters: $T_{c}=T_{c}\left(J_{\|},\, J_{\bot},\, J_{\|}^{'},\, J_{\bot}^{'},\,\eta\right)$.
Subsequently, in Section III we elaborate on influence of oxygen ordering
on $T_{c}$. We model values of in-plane phase stiffnesses as a function
of oxygen amount and determine values of model parameters for which
the YBCO phase diagram can be reproduced. In Section IV, relying on
experimental data unveiling that oxygen doping of YBCO may introduce
significant charge imbalance between CuO$_{2}$ planes and other oxygen
sites, we show that the former are underdoped, while the latter --
strongly overdoped almost in the whole region of oxygen doping in
which YBCO is superconducting. Finally, in Section V we summarize
the conclusions to be drawn from our work.

\section{Model\label{sec:Spherical-model}}

\begin{figure}
\includegraphics[%
  scale=0.35]{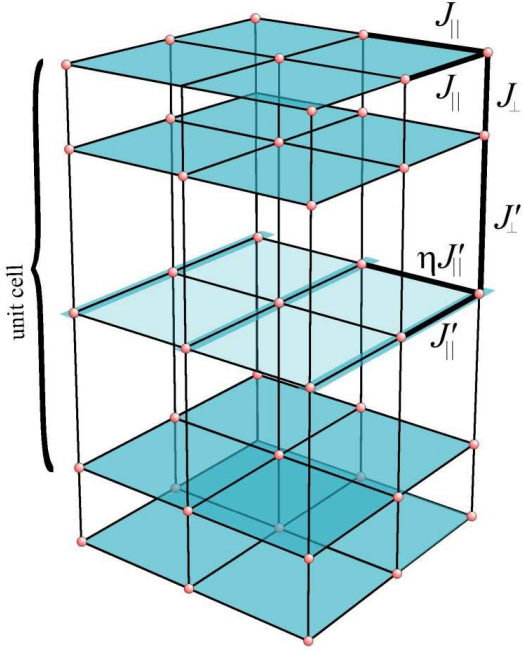}

\caption{(Color online) Structure of the YBCO superconductor. Basal plane
is in the middle of the picture with in-plane microscopic phase stiffness
$J_{\|}^{'}$, which can be strongly anisotropic when $\eta$ parameter
is small (or isotropic for $\eta=1$). Basal plane coupling with neighboring
CuO$_{2}$ planes is given by $J_{\bot}^{'}$. In-plane and inter-plane
microscopic phase stiffnesses of CuO$_{2}$ layers are $J_{\|}$ and
$J_{\bot}$, respectively. }

\label{FigModel}
\end{figure}

Since, in underdoped high-temperature superconductors, two temperature
scales of short-length pairing correlations and long-range superconducting
order seem to be well separated,\cite{EmeryKivelson} we consider
the situation, in which local superconducting pair correlations are
established and the relevant degrees of freedom are represented by
phase factors $0\le\varphi_{\ell}\left(\mathbf{r}_{i}\right)<2\pi$
placed in a lattice with nearest neighbor interactions. In our notation,
$\mathbf{r}_{i}$ numbers lattice sites within $\ell$-th $ab$ plane.
The system becomes superconducting once U(1) symmetry group governing
the $\varphi_{\ell}\left(\mathbf{r}_{i}\right)$ factors is spontaneously
broken and the non-zero value of $\left\langle e^{i\varphi_{\ell}\left(\mathbf{r}_{i}\right)}\right\rangle $
appears signaling the long-range phase order. The Hamiltonian that
we consider consists of four parts\begin{equation}
H\left[\varphi\right]=H_{\|}+H_{\bot}+H_{\|}^{'}+H_{\bot}^{'}\label{eq:Hamiltonian}\end{equation}
 containing various microscopic phase stiffnesses representing characteristic
energy scales present in YBCO (see, Fig. \ref{FigModel}) :

\begin{enumerate}
\item in-plane coupling $J_{\|}>0$ within CuO$_{2}$ layers: \begin{eqnarray}
H_{\|} & = & -J_{\|}\sum_{\ell}\sum_{i<j}\left\{ \cos\left[\varphi_{3\ell}\left(\mathbf{r}_{i}\right)-\varphi_{3\ell}\left(\mathbf{r}_{j}\right)\right]\right.\nonumber \\
 &  & +\left.\cos\left[\varphi_{3\ell+1}\left(\mathbf{r}_{i}\right)-\varphi_{3\ell+1}\left(\mathbf{r}_{j}\right)\right]\right\} ;\end{eqnarray}

\item inter-plane coupling $J_{\bot}>0$ between neighboring CuO$_{2}$
layer:\begin{eqnarray}
H_{\bot} & = & -J_{\bot}\sum_{\ell}\sum_{i<j}\cos\left[\varphi_{3\ell}\left(\mathbf{r}_{i}\right)-\varphi_{3\ell+1}\left(\mathbf{r}_{i}\right)\right];\end{eqnarray}

\item in-plane coupling $J_{\|}^{'}>0$ within basal planes, which can be
gradually turned off along $b$ direction by anisotropy parameter
$\eta\in\left[0,1\right]$ to simulate oxygen ordering:\begin{eqnarray}
H_{\|}^{'} & = & -J_{\|}^{'}\sum_{\ell}\sum_{i}\sum_{j=-1,1}\left\{ \cos\left[\varphi_{\ell}\left(\mathbf{r}_{i}\right)-\varphi_{\ell}\left(\mathbf{r}_{i}+j\hat{a}\right)\right]\right.\nonumber \\
 &  & +\left.\eta\cos\left[\varphi_{\ell}\left(\mathbf{r}_{i}\right)-\varphi_{\ell}\left(\mathbf{r}_{i}+j\hat{b}\right)\right]\right\} ;\end{eqnarray}

\item inter-plane coupling $J_{\bot}^{'}>0$ between adjacent CuO$_{2}$
layer and basal plane:\begin{eqnarray}
H_{\bot}^{'} & = & -J_{\bot}^{'}\sum_{\ell}\sum_{i<j}\left\{ \cos\left[\varphi_{3\ell+1}\left(\mathbf{r}_{i}\right)-\varphi_{3\ell+2}\left(\mathbf{r}_{i}\right)\right]\right.\nonumber \\
 &  & +\left.\cos\left[\varphi_{3\ell+2}\left(\mathbf{r}_{i}\right)-\varphi_{3\ell+3}\left(\mathbf{r}_{i}\right)\right]\right\} .\end{eqnarray}

\end{enumerate}
The indices $i,j$ go from $1$ to $N_{\|}$ being the number of sites
in a plane, $\ell=1,...,N_{\bot}/3$, where $N_{\bot}$ denotes the
number of layers and $N=N_{\|}N_{\bot}$ is the total number of sites.
The partition function of the system reads:\begin{equation}
Z=\int_{0}^{2\pi}\prod_{\ell,i}d\varphi_{\ell}\left(\mathbf{r}_{i}\right)e^{-\beta H\left[\varphi\right]},\label{eq:PartFunctionFirst}\end{equation}
where $\beta=1/k_{B}T$ with $T$ being the temperature. Introducing
two-dimensional vectors $\mathbf{S}_{\ell}\left(\mathbf{r}_{i}\right)=\left[S_{x\ell}\left(\mathbf{r}_{i}\right),\, S_{y\ell}\left(\mathbf{r}_{i}\right)\right]$
of the unit length $\mathbf{S}_{\ell}^{2}\left(\mathbf{r}_{i}\right)=S_{x\ell}^{2}\left(\mathbf{r}_{i}\right)+S_{y\ell}^{2}\left(\mathbf{r}_{i}\right)=1$
defined by $\mathbf{S}_{\ell}\left(\mathbf{r}_{i}\right)=\left[\cos\varphi_{\ell}\left(\mathbf{r}_{i}\right),\sin\varphi_{\ell}\left(\mathbf{r}_{i}\right)\right],$
the Hamiltonian can be expressed in a vector form and the partition
function written as: \begin{equation}
Z=\int_{0}^{2\pi}\left\{ \prod_{\ell,i}d^{2}\mathbf{S}_{\ell}\left(\mathbf{r}_{i}\right)\delta\left[\mathbf{S}_{\ell}^{2}\left(\mathbf{r}_{i}\right)-1\right]\right\} e^{-\beta H\left[S\right]},\label{eq:PartFunctionVector}\end{equation}
where the unit length constraint ($\mathbf{S}_{\ell}^{2}\left(\mathbf{r}_{i}\right)=1$)
is introduced by the set of Dirac-$\delta$ functions. Unfortunately,
the partition function in Eq. (\ref{eq:PartFunctionFirst}) cannot
be calculated exactly. However, the model becomes solvable, while
the rigid length constraint in Eq. (\ref{eq:PartFunctionVector})
is replaced by a weaker spherical closure relation\cite{SphericalModel}\begin{equation}
\delta\left[\mathbf{S}_{\ell}^{2}\left(\mathbf{r}_{i}\right)-1\right]\,\,\,\rightarrow\,\,\,\delta\left[\frac{1}{N}\sum_{i,\ell}\mathbf{S}_{\ell}^{2}\left(\mathbf{r}_{i}\right)-1\right].\label{SphericalConstraint}\end{equation}
Introducing different microscopic phase stiffnesses for CuO$_{2}$
and basal planes breaks translational symmetry along $c$ axis, since
the inter-plane and in-plane couplings vary with period of $3$, when
moving from one plane to another. As a result, standard way of diagonalizing
the Hamiltonian using three-dimensional Fourier transform of variables
($\mathbf{S}_{\ell}\left(\mathbf{r}_{i}\right)$ in this case) fails,
because of the lack of complete translational symmetry: To overcome
this difficulty, we implement a combination of two-dimensional Fourier
transform for in-plane vector variables \begin{equation}
\mathbf{S}_{\ell}\left(\mathbf{r}_{i}\right)=\frac{1}{N_{\|}}\sum_{\mathbf{k}}\mathbf{S}_{\mathbf{k}\ell}e^{-i\mathbf{kr}_{i}}.\end{equation}
and transfer matrix method for one-dimensional decorated structure
along $c$-axis.\cite{DecoratedLattice} This former operation diagonalizes
all terms in the Hamiltonian in Eq. (\ref{eq:Hamiltonian}) with respect
to $\mathbf{k}$, leaving the dependence on layer index $\ell$ unchanged.
Therefore, the partition function can be written in the form:\begin{eqnarray}
Z & = & \int_{-\infty}^{+\infty}\frac{d\lambda}{2\pi i}\exp\left\{ N\lambda+\frac{1}{2}\ln\int_{-\infty}^{+\infty}\prod_{\mathbf{k},\ell}d^{2}\mathbf{S}_{\mathbf{k}\ell}\right.\nonumber \\
 &  & \times\left.\exp\left[-\frac{1}{N_{\|}}\sum_{\mathbf{k},\ell,\ell'}\mathbf{S}_{\mathbf{k}\ell}A_{N_{\bot}}^{\ell\ell'}\left(\mathbf{k}\right)\mathbf{S}_{-\mathbf{k}\ell'}\right]\right\} ,\label{eq:PartFunctionFinal}\end{eqnarray}
where $A_{N_{\bot}}^{\ell\ell'}\left(\mathbf{k}\right)$ is an element
of a square $N_{\bot}\times N_{\bot}$ band matrix, appearing as a
result of non-trivial coupling structure along $c$-direction:\begin{widetext}\begin{center}

\begin{equation} \begin{array}{cc} &\hspace{1em}\\ A_{N_\bot }\left( \mathbf{k} \right) = & \left[ \begin{tabular}{cccccccc} \cline{1-3}\multicolumn{1}{|c}{$\lambda -\frac{\beta J_{\|}\left(\mathbf{k}\right) }{2}$} & $-\frac{\beta J_{\bot }^{'}}{2}$ &\multicolumn{1}{c|}{$0$}&$0$&$\cdots$&$0$&$0$&$0$ \\\multicolumn{1}{|c}{$-\frac{\beta J_{\bot }^{'}}{2}$}&$\lambda -\frac{\beta J_{\|}^{'}\left(\mathbf{k}\right) }{2}$&\multicolumn{1}{c|}{$-\frac{\beta J_{\bot }^{'}}{2}$}&$0$ &$\cdots$&$0$&$0$&$0$ \\\multicolumn{1}{|c}{$0$}&$-\frac{\beta J_{\bot }^{'} }{2}$&\multicolumn{1}{c|}{$\lambda -\frac{\beta J_{\|}\left(\mathbf{k}\right) }{2} $}&\fcolorbox[rgb]{0.9,0.9,0.9}{0.9,0.9,0.9}{$-\frac{\beta J_{\bot}}{2}$}&$\cdots$&$0$&$0$&$0$ \\\cline{1-4}$0$&$0$&$\overset{\vspace{0.1em}}{\fcolorbox[rgb]{0.9,0.9,0.9}{0.9,0.9,0.9}{$-\frac{\beta J_{\bot}}{2}$}}$&\multicolumn{1}{|c}{$\lambda -\frac{\beta J_{\|}\left(\mathbf{k}\right) }{2}$}&$\ddots$&$\vdots$&$\vdots$&$\vdots$\\$\vdots$&$\vdots$&$\vdots$&$\ddots$&\multicolumn{1}{c|}{$\lambda - \frac{\beta J_{\|}\left(\mathbf{k}\right) }{2}$}&\fcolorbox[rgb]{0.9,0.9,0.9}{0.9,0.9,0.9}{$-\frac{\beta J_{\bot}}{2}$}&$0$&$0$\\\cline{5-8}$0$&$0$&$0$&$\cdots$&$\overset{\vspace{0.1em}}{\fcolorbox[rgb]{0.9,0.9,0.9}{0.9,0.9,0.9}{$-\frac{\beta J_{\bot}}{2}$}}$&\multicolumn{1}{|c}{$\lambda -\frac{\beta J_{\|}\left(\mathbf{k}\right) }{2}$} & $-\frac{\beta J_{\bot}^{'}}{2}$ &\multicolumn{1}{c|}{$0$}\\$0$&$0$&$0$&$\cdots$&$0$&\multicolumn{1}{|c}{$-\frac{\beta J_{\bot}^{'} }{2}$}&$\lambda -\frac{\beta J_{\|}^{'}\left(\mathbf{k}\right) }{2}$&\multicolumn{1}{c|}{$-\frac{\beta J_{\bot}^{'} }{2}$}\\$0$&$0$&$0$&$\cdots$&$0$&\multicolumn{1}{|c}{$0$}&$-\frac{\beta J_{\bot}^{'} }{2}$&\multicolumn{1}{c|}{$\underset{\vspace{0.1em}}{\lambda -\frac{\beta J_{\|}\left(\mathbf{k}\right) }{2}}$} \\\cline{6-8}\end{tabular} \right] .\\ &\hspace{0.5em}\hfill \end{array} \label{MatrixAN} \end{equation}\end{center}\end{widetext}where:\begin{eqnarray}
J_{\|}\left(\mathbf{k}\right) & = & 2J_{\|}\left[\cos\left(ak_{x}\right)+\cos\left(bk_{y}\right)\right],\nonumber \\
J_{\|}^{'}\left(\mathbf{k}\right) & = & 2J_{\|}^{'}\left[\cos\left(ak_{x}\right)+\eta\cos\left(bk_{y}\right)\right]\end{eqnarray}
and $\lambda$is a Lagrange multiplier introduced by representing
the Dirac-$\delta$ function in a spectral form $\delta\left(x\right)=\int_{-\infty}^{+\infty}d\lambda/2\pi i\exp\left(-\lambda x\right)$.
The problem reduces then to evaluation of a determinant of the $A_{N_{\bot}}^{\ell\ell'}\left(\mathbf{k}\right)$
matrix (for technical details, we refer readers to Ref. {[}\onlinecite{OurPrevious}{]}).

The partition function in Eq. (\ref{eq:PartFunctionFinal}) can be
written as:\begin{equation}
Z=\int_{-\infty}^{+\infty}\frac{d\lambda}{2\pi i}\exp\left[-N\beta f\left(\lambda\right)\right].\label{eq:PartFunctFreeEnergy}\end{equation}
In the thermodynamic limit $N\rightarrow\infty$ the dominant contribution
to the integral in Eq. (\ref{eq:PartFunctFreeEnergy}) comes from
the saddle point $\lambda=\lambda_{0}$ of $f\left(\lambda\right)$:\begin{equation}
\left.\frac{\partial f\left(\lambda\right)}{\partial\lambda}\right|_{\lambda=\lambda_{0}}=0\label{eq:SaddlePoint}\end{equation}
and where $f\left(\lambda=\lambda_{0}\right)$ becomes a free energy.
In the spherical model, the emergence of the critical point is signaled
by divergence of the order parameter susceptibility:\begin{equation}
G^{-1}\left(\mathbf{k}=0\right)=0,\label{eq:ParSuscDiv}\end{equation}
where $G^{-1}\left(\mathbf{k}\right)\equiv\left\langle \mathbf{S}_{\mathbf{k},\ell}\mathbf{S}_{\mathbf{-k},\ell}\right\rangle $
and $\left\langle ...\right\rangle $ is the statistical average.
Eq. (\ref{eq:ParSuscDiv}) determines the value of the Lagrange multiplier
$\lambda$. Consequently, the free energy of the system reads:\begin{widetext}

\begin{eqnarray}
f & = & \frac{\lambda}{\beta}-\frac{1}{3\beta}\int_{-\frac{\pi}{a}}^{+\frac{\pi}{a}}\int_{-\frac{\pi}{b}}^{+\frac{\pi}{b}}\frac{dk_{x}dk_{y}}{\left(2\pi\right)^{2}/\left(ab\right)}\ln\left[\frac{1}{16}\left(\left\{ \left(\beta J_{\bot}\right)^{2}-\left[\Lambda\left(\mathbf{k}\right)\right]^{2}\right\} \Lambda^{'}\left(\mathbf{k}\right)+2\left(\beta J_{\bot}^{'}\right)^{2}\Lambda\left(\mathbf{k}\right)\right)\right.\nonumber \\
 &  & \left.+\sqrt{\left\{ \left(\beta J_{\bot}\right)^{2}-\left[\Lambda\left(\mathbf{k}\right)\right]^{2}\right\} \left\{ \left(\beta J_{\bot}\right)^{2}\left[\Lambda^{'}\left(\mathbf{k}\right)\right]^{2}-\left[2\left(\beta J_{\bot}^{'}\right)^{2}-\Lambda\left(\mathbf{k}\right)\Lambda^{'}\left(\mathbf{k}\right)\right]^{2}\right\} }\right],\end{eqnarray}
where functions $\Lambda\left(\mathbf{k}\right)=\beta J_{\|}\left(\mathbf{k}\right)-2\lambda_{0}$
and $\Lambda^{'}\left(\mathbf{k}\right)=\beta J_{\|}^{'}\left(\mathbf{k}\right)-2\lambda_{0}$.
The Lagrange multiplier reads:\begin{equation}
\lambda_{0}=\frac{\beta}{4}\left\{ 4J_{\|}+2J_{\|}^{'}\left(1+\eta\right)+J_{\bot}+\sqrt{8\left(J_{\bot}^{'}\right)^{2}+\left[J_{\bot}+4J_{\|}+2J_{\|}^{'}\left(1-\eta\right)\right]^{2}}\right\} .\end{equation}
 Finally, the Eq. (\ref{eq:SaddlePoint}) leads to expression for
the critical temperature:\begin{equation}
\beta_{c}=\frac{2}{3}\int_{-1}^{+1}d\xi\int_{-1}^{+1}d\zeta\rho\left(\xi\right)\rho\left(\zeta\right)\frac{2\left(J_{\bot}^{'}\right)^{2}+\left(J_{\bot}\right)^{2}-4\alpha\left(\alpha+2\alpha^{'}\right)}{\sqrt{\left\{ \left[\left(J_{\bot}^{'}\right)^{2}-2\alpha\alpha^{'}\right]^{2}-\left(J_{\bot}\alpha^{'}\right)^{2}\right\} \left[\left(2\alpha\right)^{2}-J_{\bot}^{2}\right]}},\label{eq:CriticalTemperature}\end{equation}
\end{widetext}where $\alpha\equiv\lambda_{0}/\beta-J_{\|}\left(\xi+\zeta\right)$,
$\alpha^{'}\equiv\lambda_{0}/\beta-J_{\|}^{'}\left(\xi+\eta\zeta\right)$
and $\rho\left(\varepsilon\right)$ is a density of states of the
chain (one-dimensional) lattice given by:\begin{equation}
\rho\left(\varepsilon\right)=\frac{1}{\pi}\frac{1}{\sqrt{1-\varepsilon^{2}}}\Theta\left(1-\left|\varepsilon\right|\right),\end{equation}
 and $\Theta\left(x\right)$ is the unit-step function.

\section{Influence of oxygen ordering on the critical temperature}

\begin{figure}
\includegraphics[%
  scale=0.65]{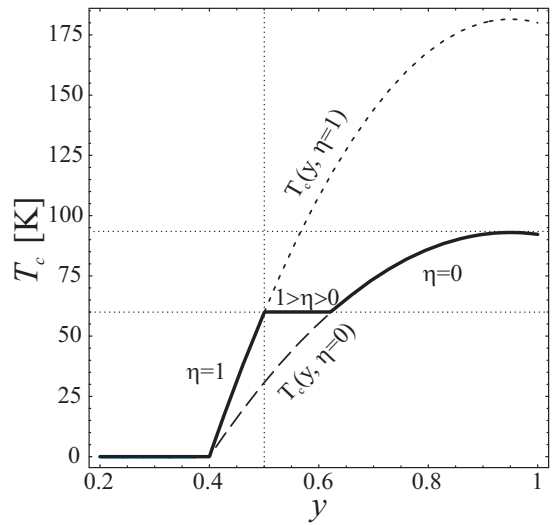}

\caption{Construction of YBCO phase diagram within presented model: curves
for $T_{c}\left(y,\,\eta\right)$ for disordered ($\eta=1$) and fully
ordered ($\eta=0$) oxygen set limits for the actual phase diagram.
For low dopings ($y<0.5$) interactions in the basal planes are isotropic
and $T_{c}\left(y\right)$ dependence falls on $\eta=1$ curve. On
the other hand, for $y=1$, all added oxygen atoms form chains and
the critical temperature reaches the other limit of $\eta=0$. Plateau
region is a crossover between those two regimes in which variable
value of $\eta$ keeps $T_{c}$ conA. J. Leggett}

\caption{Affiliation: Department of Physics, University of Illinois at Urbana-Champaign,
Urbana, Illinois 61801, USAstant. \label{cap:Construction-of-YBCO}}
\end{figure}

The result in Eq. (\ref{eq:CriticalTemperature}) provides us with
a tool to analyze the influence of anisotropy and characteristic energy
scales present in YBCO on the critical temperature. However, in order
to describe YBCO phase diagram, it is necessary to connect the model
parameters with amount of oxygen doping. In our previous paper we
have proposed a phenomenological dependence of in-plane microscopic
phase stiffness of CuO$_{2}$ plane on charge (hole) concentration,
which appeared to successfully describe properties of superconducting
homologous series:\cite{OurPrevious}\begin{equation}
J_{\|}\left(\delta\right)=\left\{ \begin{array}{l}
J_{\|}\left[1-\frac{1}{0.01}\left(\delta-0.15\right)^{2}\right]\,\,\mathrm{for}\,\,0.05<\delta<0.025\\
0\,\,\,\mathrm{for}\,\,\delta\le0.05\,\,\mathrm{or\,\,\delta\ge0.25}.\end{array}\right.\label{eq:JpFromOurPrevious}\end{equation}
 To adapt it to the present model it is necessary to relate oxygen
amount to charge concentration in CuO$_{2}$ planes, since YBCO phase
diagram is presented as an oxygen doping function of temperature.
This relation was experimentally determined by Tallon, et at. and
found out to be roughly linear in the superconducting region (charge
concentration changes from $\delta=0.05$ for $y=0.4$ to $\delta=0.17$
for $y=1$).\cite{GenericBehaviorYBCO} Consequently, the in-plane
phase stiffness from Eq. (\ref{eq:JpFromOurPrevious}) expressed as
a function of oxygen amount reads $J_{\|}\left(y\right)=J_{\|}g\left(y\right)$,
where:\begin{equation}
g\left(y\right)=\left\{ \begin{array}{l}
1-\frac{1}{0.55^{2}}\left(y-0.95\right)^{2}\,\,\mathrm{for}\,\,0.4\le y\le1\\
0\,\,\,\mathrm{for}\,\,\mathrm{y\le0.4}.\end{array}\right.\label{eq:g_y}\end{equation}
Although, it is usually stated in the literature, T-O phase transition
and emergence of superconductivity in YBCO are not simultaneous in
terms of doping.\cite{OTPhaseTransition} Thus, we suggest the following
scenario describing the phase diagram of YBCO (see, Fig. \ref{cap:Construction-of-YBCO}):
with increasing oxygen doping (starting from $y=0$), the onset of
superconductivity is reached ($y=0.4$). The critical temperature
is rising up to a point, in which T-O phase transition occurs ($y=0.5$).
For higher dopings, oxygen chains start to form. That results in interactions
in basal planes becoming more one-dimensional, thus phase fluctuations
rise significantly and enough to keep the critical temperature constant.
Further, while all the oxygen is ordered in chains, the critical temperature
roughly follows in-plane phase stiffness $J_{\|}$ dependence on doping
reaching its maximum value of $93\mathrm{K}$ for $y=0.95$ and dropping
slightly later.

\begin{figure}
\includegraphics[%
  scale=0.7]{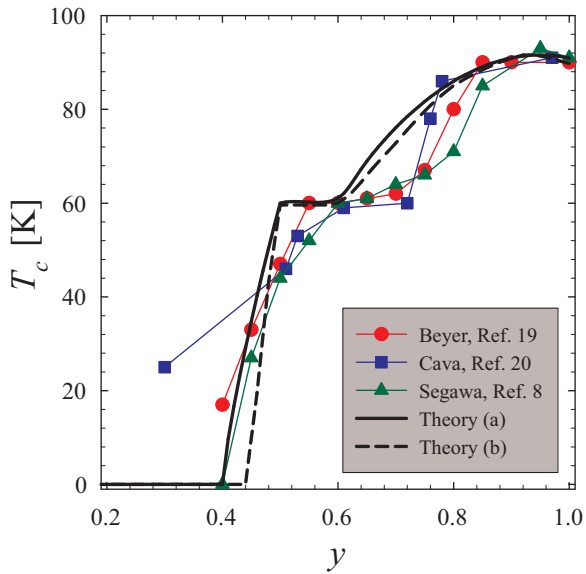}

\caption{(Color online) Comparison of experimental phase diagrams of YBCO
with results of the presented model for a) linear and b) non-linear
(more exact) approximation of data of Tallon, et al. (see, Ref \onlinecite{GenericBehaviorYBCO}).\label{cap:PhaseDiagramOrder}}
\end{figure}

In order to realize this scenario within the presented model it is
necessary to fix values of model parameters $J_{\|}$, $J_{\|}^{'}$,
$J_{\bot}$, $J_{\bot}^{'}$ and find a dependence of $\eta$ on doping
$y$, which would results in constant critical temperature in the
plateau region of the phase diagram. We assume that doping dependence
of $J_{\|}$ and $J_{\|}^{'}$ are given by Eq. (e\ref{eq:g_y}),
and, for simplicity, $J_{\bot}\equiv J_{\bot}^{'}$. Experimental
data for anisotropy of penetration depth in YBCO provide us with ratio
of $J_{\bot}/J_{\|}=\lambda_{ab}^{2}/\lambda_{c}^{2}\simeq100$.\cite{PenetrationDepth}
Furthermore, values of $J_{\|}$ and $J_{\|}^{'}$ can be deduced
from the ratio of expressions for the critical temperatures (see,
Fig. \ref{cap:Construction-of-YBCO}) for plateau ($60K$) and optimum
doping ($93K$):\begin{equation}
\frac{T_{c}\left(J_{\|,}\, J_{\|}^{'},\, y=0.5,\,\eta=1\right)}{T_{c}\left(J_{\|,}\, J_{\|}^{'},\, y=0.95,\,\eta=0\right)}=\frac{60}{93}.\end{equation}
Finally, dependence of oxygen ordering on doping $\eta\left(\delta\right)$
can be found numerically for calculated values of model parameters
($J_{\|}=19.32\textrm{meV}$, $J_{\|}^{'}=33.81\mathrm{meV}$, $J_{\bot}=J_{\bot}^{'}=0.58\mathrm{meV}$)
and it can by approximated by the expression:\begin{equation}
\eta\left(\delta\right)=\frac{1}{3}\left[\left(2-2\delta\right)^{15.2}+\left(2-2\delta\right)^{7.9}+\left(2-2\delta\right)^{3.6}\right].\end{equation}
The resulting phase diagram is presented in Fig. \ref{cap:PhaseDiagramOrder}
(thick solid line) along with experimental results, for comparison.
It is clearly seen that the procedure reproduces characteristic features
of YBCO phase diagram: 60-K plateau, 93K maximum critical temperature
with small overdoped region. However, trying to fit experimental data
better, we have performed more exact (non-linear) approximation of
data from Ref. {[}\onlinecite{GenericBehaviorYBCO}{]} that lead to
doping dependence of in-plane phase stiffness $J_{\|}\left(y\right)$.
Corresponding phase diagram is also presented in Fig. \ref{cap:PhaseDiagramOrder}
as thick dashed line. It does not provide any new quality increasing
the width of 60-K plateau only very slightly, thus it is reasonable
to use simpler form of $J_{\|}\left(y\right)$ dependence, as in Eq.
(\ref{eq:g_y}). It can be also noticed that the model predicts 60-K
plateau to be a little bit narrower and moved toward lower doping
region than experimental results show. It can be argued that since
it is very hard to control the amount of oxygen in YBCO precisely,
some measurements, especially based on multigrain powder samples,
can be inaccurate. However newer results are based on un-twinned single
crystals and seem to be reliable, also because of good consistence
among different research groups.

\begin{figure}
\includegraphics[%
  scale=0.7]{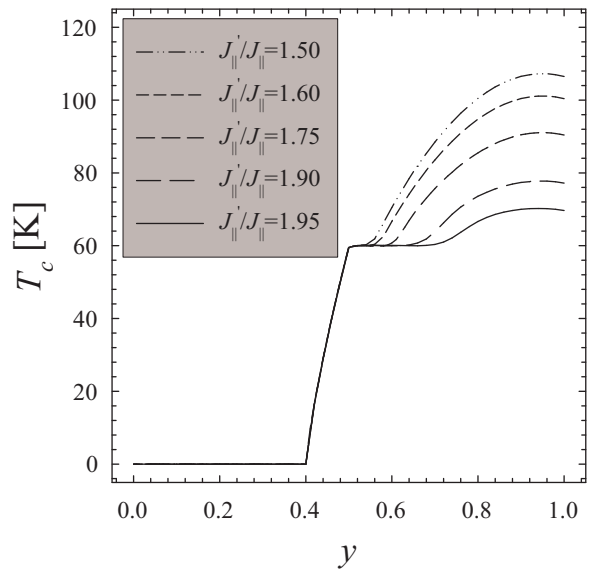}

\caption{Comparison of phase diagrams for various ratios of $J_{\|}^{'}/J_{\|}$.
Plots are normalized to have the critical temperature of the plateau
region equal to $60\mathrm{K}$.\label{cap:Comp-phase-diagrams-JJ}}
\end{figure}

Unfortunately, the most serious flaw in the presented scenario are
the actual values of model parameters required to reproduce experimental
phase diagram. It is necessary for basal plane in-plane microscopic
phase stiffness $J_{\|}^{'}$ to be almost two times bigger than CuO$_{2}$
planes coupling $J_{\|}$. For more reasonable values $J_{\|}>J_{\|}^{'},\, J_{\bot},\, J_{\bot}^{'}$,
the influence of anisotropy in basal planes is almost negligible.
For example for $J_{||}=J_{\|}^{'}=J$, $J_{\bot}^{'}=0.1J$ and $J_{\bot}=0.01J$
with $J=19.3\mathrm{meV}$, the ratio of critical temperatures for
systems with anisotropic ($\eta=0$) and isotropic ($\eta=1$) interactions
in basal planes reads:\begin{equation}
\frac{T_{c}\left(\eta=0\right)}{T_{c}\left(\eta=1\right)}=0.949.\end{equation}
Similarly, for $J_{||}=J_{\|}^{'}=J$, $J_{\bot}^{'}=0.1J$ and $J_{||}=J$,
$J_{||}^{'}=J_{\bot}^{'}=0.5J$ and $J_{\bot}=0.1J$ the ratio is
$0.983$. The influence of basal plane anisotropy is too small to
noticeably change the critical temperature until in-plane phase stiffness
$J_{\|}^{'}$ is increased to be higher than $J_{\|}$. This, however,
does not seem to be reasonable from physical point of view. Evolution
of the phase diagram with changing ratio of $J_{\|}^{'}/J_{\|}$ is
presented in Fig. \ref{cap:Comp-phase-diagrams-JJ}. When the ratio
is being decreased, the plateau region is disappearing. To summarize,
we find it rather unlikely that oxygen ordering into chains alone
can explain the existence of 60-K plateau. 

\section{Influence of doping imbalance on the critical temperature}

YBCO can be doped not only by adding the oxygen atoms, but also by
substituting three-valent Yttrium atoms with two-valent Calciums.
This introduces holes directly to CuO$_{2}$ planes leaving apical
and chain sites untouched:\cite{Apical_oxygen} even though the charge
concentration within copper-oxide planes is high enough, lack of doping
of apical sites results in interplane coupling being small and the
onset of superconductivity is not reached. This suggests that charge
concentration within apical sites along with basal plane sites and
CuO$_{2}$ layers can be significantly different. Similarly, as Yttrium
substitution can lead to introduction of holes exclusively into CuO$_{2}$
planes, one can imagine that increasing of oxygen amounts changes
charge concentration primarily in the chain regions and only some
of the charges are transported to the copper-oxide layers. In fact,
this scenario has been confirmed by means of site-specific X-ray absorption
spectroscopy.\cite{Apical_oxygen} Because overdoping of some regions
of YBCO could provide a factor decreasing the critical temperature
with increasing oxygen doping and thus leading to the 60-K plateau,
we want to investigate such a possibility within presented model.

\begin{figure}
\includegraphics[%
  scale=0.35]{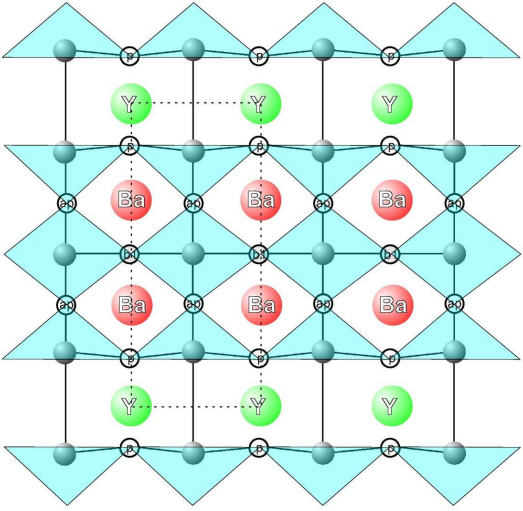}

\caption{(Color online) YBa$_{2}$Cu$_{3}$O$_{7}$ structure with copper-oxide
chains fully formed. \label{cap:YBCO_chains}}
\end{figure}

First, we want to emphasize the high significance of apical site doping.
Analyzing a projection of the YBCO structure on $a-c$ plane (presented
in Fig. \ref{cap:YBCO_chains}), it can be noticed that a set of chains
differs from full copper-oxide layer only by a lack of oxygens between
CuO$_{2}$ planes. Thus one can expect that the microscopic phase
stiffnesses $J_{\bot}^{'}$ related to apical sites and the one within
basal planes $J_{\|}^{'}$ are of the order of CuO$_{2}$ in-plane
phase stiffness $J_{\|}$. This suggests it is reasonable to investigate
the role of doping of apical and basal sites on the phase diagram
of YBCO. To investigate this , we want to determine values of $J_{\|}^{'}$
and $J_{\bot}^{'}$ that would result in $T_{c}\left(y\right)$ dependence
observed experimentally. For clarity, we assume that oxygen is fully
ordered in chains to study the effect of charge imbalance alone. It
is also necessary to model an influence of the number of vacancies
in basal planes for $y<1$ on the in-plane phase stiffness $J_{\|}^{'}$.
This relation is unfortunately not obvious and it is hard to find
any hint about it specific form. Thus, we start with a simple linear
dependence $h\left(y\right)=y$, so : $J_{\| effective}^{'}\left(y\right)=y^{\gamma}J_{\|}^{'}\left(y\right)$
with $\gamma=1$.\cite{OxygenVacanciesComment} Later, we choose different
value of $\gamma=3/2$ and notice that although specific values of
$J_{\|}^{'}$ and $J_{\bot}^{'}$ change, the qualitative results
are the same and in reasonable agreement with experimental data.

\begin{figure}
\includegraphics[%
  scale=0.7]{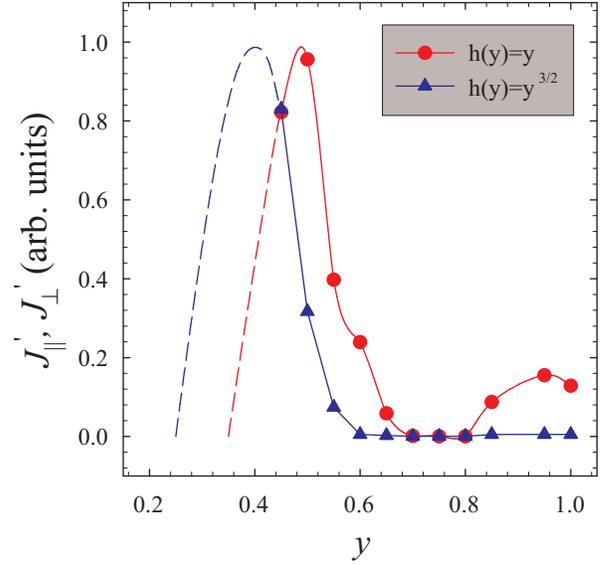}

\caption{(Color online) Calculated dependence of $J_{\|}^{'}$ and $J_{\bot}^{'}$
on oxygen amount for various factors due to number of oxygen vacancies
in copper-oxide chains. Dependences are normalized to have maximum
value equal to 1. Dashed lines are guides for eye extending dependences
beyond region where calculation was possible.\label{cap:JrJp}}
\end{figure}

\begin{figure}
\includegraphics[%
  scale=0.7]{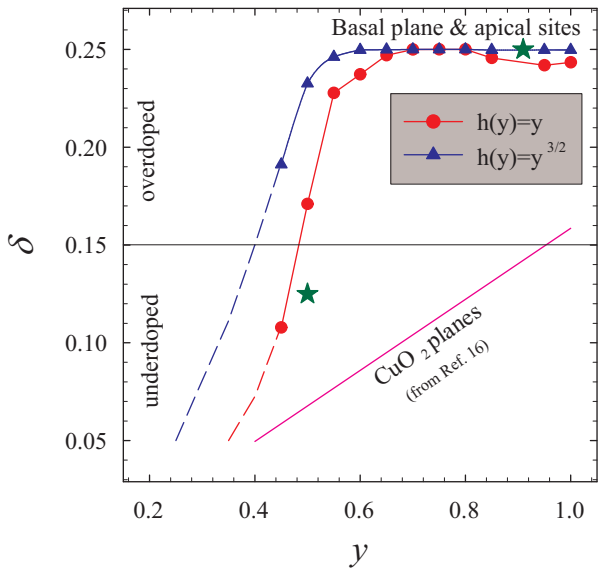}

\caption{(Color online) Charge concentration within basal planes, apical sites
and in CuO$_{2}$ planes. Stars denote experimental values of hole
doping in basal planes and apical sites (from Ref. {[}\onlinecite{Apical_oxygen}{]}).
Dashed lines are guides for eye extending dependences beyond region
where calculation was possible.\label{cap:rho_y}}
\end{figure}

Our procedure is as follows: we assume that anisotropy ratio between
in-plane and inter-plane coupling among CuO$_{2}$ planes is equal
to $J_{\bot}/J_{\|}=0.0136$,\cite{OurPrevious} $\eta$ parameter
is equal to $0$ (oxygen in basal planes is ordered along $a$ direction)
and values of $J_{\|}^{'}$ and $J_{\bot}^{'}$ are equal and of the
order of $J_{\|}$ (however $J_{\|}^{'}$ is modified by chosen factor
$h\left(y\right)=y^{\gamma}$). Providing values of $J_{\|}$, $J_{\bot}$
and $\eta$ into Eq. (\ref{eq:CriticalTemperature}) along with experimentally
obtained critical temperatures based on resistivity measurements from
Ref. {[}\onlinecite{YBCO_transport}{]}, we determine doping dependence
of basal and apical microscopic phase stiffnesses. The results are
presented in Fig. \ref{cap:JrJp}. Using expression in Eq. (\ref{eq:JpFromOurPrevious})
it is also possible to calculate doping dependence of corresponding
charge concentration (see, Fig. \ref{cap:rho_y}). As it is apparent,
almost in the whole region of oxygen doping in which YBCO is superconducting,
basal planes along with apical sites are overdoped while \emph{simultaneously}
CuO$_{2}$ planes are underdoped. While oxygen content is increased,
this provides two counteracting factors, which may lead to rise of
60K plateau. Although specific relation of $J_{\|}^{'}$ and $J_{\bot}^{'}$
on $y$ is dependent on assumed factor $h\left(y\right)=y^{\gamma}$,
the results are qualitatively similar and in reasonable agreement
with experimental data (see stars, in Fig. \ref{cap:rho_y}).


\section{Summary and conclusions}

We have considered a model of YBa$_{2}$Cu$_{3}$O$_{6+y}$ to study
the influence of oxygen ordering and doping imbalance on the critical
temperature $T_{c}\left(y\right)$ and to elucidate a possible origin
of well-known feature of YBCO phase diagram: the 60-K plateau. Motivated
by the experimental evidence that the ordering of the phase degrees
of freedom is responsible for the emergence of the superconducting
state with long-range order, we focus on the ``phase only'' description
of the high-temperature superconducting system. In our approach, the
vanishing of the superconductivity with underdoping can be understood
by the reduction of the in-plane phase stiffnesses and can be linked
to the manifestation of Mott physics (no double occupancy due to the
large Coulomb on-site repulsion) that leads to the loss of long-range
phase coherence while moving toward half-filled limit ($y=0$). For
doping $y=0$, the fixed electron number implies large fluctuations
in the conjugate phase variable, which naturally translates into the
reduction of the microscopic in-plane phase stiffnesses and destruction
of the superconducting long-range order in this limit. In the opposite
region of large $y$, the onset of a pair-breaking effect (at the
pseudogap temperature $T^{*}$) can deplete the microscopic phase
stiffnesses, thus reducing the critical temperature. 

To this end, we have utilized a three-dimensional semi microscopic
XY model with two-component vectors that involve phase variables and
adjustable parameters representing microscopic phase stiffnesses and
an anisotropy parameter. The model fully implements complicated energy
scales present in YBCO also allowing for strong anisotropy within
basal planes in order to simulate oxygen ordering. Applying spherical
closure relation we have solved the phase XY model with the help of
transfer matrix method and calculated $T_{c}$ for chosen system parameters.
Furthermore, by making a physically justified assumption regarding
the doping dependence of the microscopic phase stiffnesses we are
able to recreate the phase diagram of YBCO and investigate the influence
of oxygen ordering and doping imbalance on its shape. We determine
that characteristic features of the phase diagram, i.e. 60-K plateau,
can be recreated by effects of oxygen ordering. However, the specific
values of the model parameters for which this result is obtained seem
to be a little bit hard to justify. Furthermore, relying on experimental
data unveiling that oxygen doping of YBCO may introduce significant
charge imbalance between CuO$_{2}$ planes and other oxygen sites,
we show that the former are underdoped, while the latter -- strongly
overdoped almost in the whole region of oxygen doping in which YBCO
is superconducting. Increasing of the oxygen content provides then
a natural mechanism of two counter acting factors that leads to emergence
of 60-K plateau. Additionally, our result can provide an important
contribution to solve a controversy of the symmetry of YBCO order
parameter, for which various experiments give contradictory answers
suggesting $d$ or $s$-wave (although the fact that YBCO is orthorhombic
should also lead to order parameter being a mixture of $s$ and $d$-wave).
\cite{OrderParam1} Furthermore, measurements of the complex conductivity
of high quality YBCO crystals show a third peak in the normal conductivity
at 80K along with enhanced pair conductivity below $\sim60\mathrm{K}$.\cite{OrderParam2}
Authors show that a single $d$-wave order parameter is insufficient
to describe the data and successfully consider two-component model
of superconductivity in YBCO. They claim that it would be tempting
to assign the two superconducting components with the associated condensates
residing on CuO$_{2}$ planes and chains, respectively, however they
do not see any justification of such situation. Thus, our result can
provide a natural answer for plausibility of such a scenario, in which
$s$-wave component comes from overdoped chains and $d$-wave one
-- from underdoped CuO$_{2}$ planes. 

\begin{acknowledgments}
T. A. Z. would like to thank The Foundation for Polish Science for
supporting his stay at the University of Illinois within Foreign Postdoc
Fellowships program. T. A. Z. would also like to thank Prof. Leggett
for hospitality during the stay, fruitful discussions and invaluable
comments to the manuscript.
\end{acknowledgments}


\begin{thebibliography}{10}
\bibitem{OxygenVacanciesComment}In order to find exact influence of number of oxygen vacancies in
chains on $J_{\|}^{'}$ it would be necessary to calculate doping-dependent
anisotropic penetration depth and fine tune it to experimental results.
\end{thebibliography}
\end{document}